\documentclass[12pt]{article}

\usepackage{amssymb}

\let\a=\alpha   \let\b=\beta   \let\g=\gamma   \let\d=\delta
\let\e=\epsilon    \let\h=\eta     \let\q=\theta
      \let\l=\lambda  \let\m=\mu
\let\n=\nu           \let\p=\pi      
\let\s=\sigma   \let\t=\tau     

     \let\L=\Lambda

\newcommand{\Realint}{\mathbb R}

\newcommand{\be}{\begin{equation}}
\newcommand{\ee}{\end{equation}}
\newcommand{\bea}{\begin{eqnarray}}
\newcommand{\eea}{\end{eqnarray}}
\newcommand{\ba}{\begin{array}}
\newcommand{\ea}{\end{array}}

\usepackage{graphics}
\begin{document}
%
%
\begin{titlepage}
\rightline{hep-th/0402105}
\rightline{CPHT-RR 111.1203}
\vskip 2cm
\centerline{{\large\bf Anomalous U(1)s masses in
non-supersymmetric}}
\vskip 0.3cm
\centerline{{\large\bf open string vacua}}
\vskip 1cm
\centerline{P. Anastasopoulos\footnote{panasta@physics.uoc.gr}}
\vskip 1cm
\centerline{Department of Physics, University of Crete,}
\vskip 0.2cm
\centerline{71003 Heraklion, GREECE}
\vskip 0.2cm
\centerline{and}
\vskip 0.2cm
\centerline{Laboratoire de Physique Th{\'e}orique Ecole
Polytechnique}
\centerline{91128, Palaiseau, FRANCE.}

\begin{abstract}

Anomalous U(1)s are omnipresent in realizations of the Standard
Model using D-branes. Such models are typically
non-supersymmetric, and the anomalous U(1) masses are potentially
relevant for experiment. In this paper, the string calculation of
anomalous U(1) masses (hep-th/0204153) is extended to
non-supersymmetric orientifolds.

\end{abstract}
\end{titlepage}

\section{Introduction}

Recently, many attempts have been made in order to embed the
Standard Model in open string theory, with partial success
\cite{Aldazabal:2000dg, Ibanez:2001nd, Blumenhagen:2000ea,
Cvetic:2002qa, Bailin:2000kd, Kokorelis:2002ip,
Antoniadis:2000en}. In such a context the Standard Model particles
are open string states attached on (different) stacks of D-branes.
$N$ coincident D-branes away from an orientifold plane typically
generate a unitary group $U(N)$. Therefore, every $U$-factor in
the gauge group supplies the model with extra abelian gauge
fields\footnote{There are cases where we can also have $SO(n)$ or
$Sp(n)$ gauge factors. However, $SU(3)$ can be minimally embedded
only in $U(3)$ and in non-minimal cases (bigger gauge groups that
are then broken by projections to those of the Standard Model),
they leave also other potentially anomalous U(1)s.}.

Such $U(1)$ fields have generically 4D anomalies. The anomalies
are cancelled via the Green-Schwarz mechanism \cite{Green:sg,
Sagnotti:1992qw, Ibanez:1998qp} where a scalar axionic field
(zero-form, or its dual two-form) is responsible for the anomaly
cancellation. This mechanism gives a mass to the anomalous $U(1)$
fields and breaks the associated gauge symmetry. The masses of the
anomalous $U(1)$s are typically of order of the string scale but
in open string theory they can be also much lighter
\cite{Scrucca:2002is, Antoniadis:2002cs}.
If the string scale is around a few TeV, observation of such
anomalous $U(1)$ gauge bosons becomes a realistic possibility
\cite{Kiritsis:2002aj}.

As it has been shown in \cite{Antoniadis:2002cs}, we can compute
the general mass formulae of the anomalous $U(1)$s in
supersymmetric models by evaluating the ultraviolet tadpole of the
one-loop open string diagram with the insertion of two gauge
bosons on different boundaries. In this limit, the diagrams of the
annulus with both gauge bosons in the same boundary and the
M\"obius strip do not contribute when vacua have cancelled
tadpoles. Mass formulae were provided for $N=1$ and $N=2$
supersymmetric orientifolds.

In this paper we are interested in the masses of the anomalous
$U(1)$s in non-supersymmetric models since such are the models
that will eventually represent the low energy physics of the
Standard Model. In particular, intersecting-brane realizations of
the Standard Model are generically non-supersymmetric.
We calculate the mass formulae using the "background field method"
and find that they are the same as the supersymmetric ones when we
have cancellation of all tadpoles. In cases where NSNS tadpoles do
not vanish, there are extra contributions proportional to the
non-vanishing tadpole terms.

The formulae are valid even if we add Wilson lines that move the
branes away from the fixed points. The Wilson lines generically
break the gauge group and they will affect the masses of the
anomalous $U(1)$s through the traces of the model dependent $\g$
matrices.

The formulae, are applied to a $Z_2$ non-supersymmetric
orientifold model, with RR and NSNS tadpoles to be cancelled,
where supersymmetry is broken by a Scherk-Schwarz deformation
\cite{Anastasopoulos:2003ha}.

This ultraviolet mass is not the only source for the mass of
anomalous $U(1)$s. In Standard Model realizations, the Higgs is
necessarily charged under one of the anomalous $U(1)$s. As it was
described in \cite{Kiritsis:2003mc}, the Higgs contribution to the
mass of these $U(1)s$ is $g_A \sqrt{M^2+e_H^2 \langle H
\rangle^2}$ where $g_A$ the gauge coupling of the anomalous $U(1)$
and $e_H$ the $U(1)$ charge of the Higgs.
The Higgs contribution to the U(1) mass can be obtained from the
effective field theory unlike the ultraviolet mass we calculate
here which can only calculated in string theory.

The paper is organized as follows. In Section 2, we evaluate the
general mass of the anomalous $U(1)$s using the background-field
method. In Section 3, we review the non-supersymmetric $Z_2$
orientifold with a Scherk-Schwartz deformation, and we use the
results of the previous section to calculate the anomalous U(1)
masses.

\section{Computing with the background-field method}

Our purpose is to evaluate the bare masses of the anomalous $U(1)$
which appear in the one-loop amplitudes with boundaries where two
gauge fields are inserted \cite{Antoniadis:2002cs}.
Here we will use another technique which is based on turning on a
magnetic field on the D-branes and pick out the second order terms
to this magnetic field. This method is called "the
background-field method" \cite{Bachas:bh}. We turn on different
magnetic fields $B_a$ in every stack of branes, longitudinal to
$x^1$, a  non-compact dimension,
\be F^a_{23} = B_a Q_a \ , \label{MagneticField} \ee
where $Q_a$ are the $U(1)_a$ generators from every stack of
branes. The effect of the magnetic field on the open-string
spectrum is to shift the oscillator frequencies of the string
non-compact $x^2+ix^3$ coordinate by an amount $\e_a$:
\be \e_a = {1\over \p} [\arctan (\p q^a_i B_a) +  \arctan (\p
q^a_j B_a)] \ , \label{Shift} \ee
where $q^a_i$, $q^a_j$ are the $U(1)_a$ charges of the $i,j$
endpoints. The Chan-Paton states $\l_{ij}$ that describe the
endpoint $i,j$ of the open string, are the generators of gauge
group that remains after the orientifold construction.
Diagonalizing these matrixes, we can replace the $Q_i$ with
$\l_{ii}$.

The expansion of the one-loop vacuum energy is:
\bea \L(B) &=& {1 \over 2} \left( {\cal T}+{\cal K} + {\cal A}(B)
+ {\cal M}(B) \right) =
\L_0 
+ {1 \over 2} \left({B\over 2\p}\right)^2 \L_2 + \cdots \ ,
\label{oneloopvacuum} \eea
where $B$ one of the different magnetic fields.
%
%
%
Generically, it appears a linear to $B$ term that is a pour
tadpole and it is coming from the RR sector. This term vanishes
when we have tadpole cancellation.
The quadratic term in the background field contains a lot of
information. In the IR limit, we have a logarithmic divergence
whose coefficient is the $\b$-function. The UV limit provides the
mass-term of the anomalous gauge bosons. The finite part of this
term is the threshold correction in the gauge couplings
\cite{Bachas:bh}. The annulus amplitude in the $Z_N$ type I
orientifolds (without the magnetic field) can be written as:
\be {\cal A}^{ab} =-{1\over 2N} \sum_{k=1}^{N-1} \int_0^\infty \,
{dt\over t} {\cal A}^{ab}_k(q) \ , \label{BGNDanm} \ee
where $a,b$ the different kind of $D$-branes at the ends of the
open strings. The ${\cal A}^{ab}_k$ is the contribution of the
$k$th sector:
\bea {\cal A}^{ab}_k &=& {1 \over 4 \p^4 t^2} {\rm Tr} [\g^k_a]
{\rm Tr} [\g^k_b] \sum_{\a,\b=0,1} \h^{\a\b}
{\vartheta[^\a_\b]\over \h^3}
Z^{ab}_{int,k}\left[^{\a}_{\b}\right]|_{\cal A}~.
\label{BGNDannulus} \eea
Similarly, we can exchange $\cal A$ with $\cal M$ in
(\ref{BGNDanm}) to have an analogous expression for the M\"obius
strip. The ${\cal M}^{a}_k$ is given by:
\bea
{\cal M}^{a}_k &=& -{1 \over 4 \p^4 t^2} {\rm Tr}[\g_a^{2k}]
\sum_{\a,\b=0,1} \h^{\a\b} {\vartheta[^\a_\b]\over \h^3}
Z^a_{int,k}\left[^{\a}_{\b}\right]|_{\cal M}. \label{BGNDmobius}
\eea
In the presence of the background magnetic field $B_a$, the above
amplitudes become:
\bea {\cal A}^{ab}_k(B) &=& {i \over 4 \p^3 t} {\rm Tr} \left[(B_a
\l_a \g^k_a \otimes \g^k_b + \g^k_a \otimes B_b \l_b \g^k_b)
\sum_{\a\b} \h^{\a\b} {\vartheta[^\a_\b ]({i \e t \over 2})\over
\vartheta[^1_1]({i \e t \over 2})} \right]
Z^{ab}_{int,k}\left[^{\a}_{\b}\right]|_{\cal A}, \nonumber \\
{\cal M}^{a}_k(B) &=& -{i \over 2 \p^3 t} {\rm Tr} \left[ B_a \l_a
\g_a^{2k} \sum_{\a\b} \h^{\a\b} {\vartheta[^\a_\b]({i \e t \over
2})\over \vartheta[^1_1]({i \e t \over 2})} \right]
Z^a_{int,k}\left[^{\a}_{\b}\right]|_{\cal M}~, \label{BGNDon} \eea
Notice that the only differences from (\ref{BGNDannulus},
\ref{BGNDmobius}) are in the contribution of the non-compact part
of the partition functions. This is expected since the presence of
the magnetic fields affect only the $x^2, x^3$ coordinates.
Therefore, the expressions (\ref{BGNDon}) are valid for all kinds
of orientifold models.

Since we are interested in the quadratic $B^2$ terms of the above
amplitudes, we expand the above formulae to quadratic order in the
background field\footnote{Where the normalized expansion is ${\cal
A} \equiv {\cal A}_0 + {B\over 2\p} {\cal A}_1+\left({B\over
2\p}\right)^2 {\cal A}_2 + \cdots$. Similarly for ${\cal M}$.},
using the following Taylor expansions:
\be \e  \simeq \left\{ \ba{ccccc} B_a \l_a \otimes 1 + 1 \otimes
B_b \l_b & \quad & \textrm{in} & \quad & {\cal A}^{ab} ,\\ 2 B_a
\l_a \ \ \ \ \ \ \ \ \ \ \ \ \ \ \ \ \ \ & \quad & \textrm{in} &
\quad & {\cal M}^{a} . \ea \right. \label{Shift-2}\ee
The zero-order $B$ terms are the amplitudes in the absence of the
magnetic field (\ref{BGNDannulus}, \ref{BGNDmobius}). These
expressions give the tadpole cancellation conditions in virtue of
the UV divergences.
%
%
The linear to $B$ terms appear from the $a=b=1$ sector in
(\ref{BGNDon}). This is a pour tadpole and vanishes when we have
tadpole cancellation. Therefore, it does not affect higher order
in $B$ amplitudes.
%
%
The second order-terms on $B$ are:
\bea {\cal A}_{2,k}^{ab}
&=& \p^2 i~
\Big[{\rm Tr} [\l_a^2 \g_a^k]{\rm Tr} [\g_b^k]+ {\rm Tr}
[\g_a^k]{\rm Tr} [\l_b^2 \g_b^k] 
+2{\rm Tr} [\l_a\g_a^k]{\rm Tr} [\l_b\g_b^k]\Big]
F^{ab}_k|_{\cal A} \label{BGNDannulusLAST} \\
{\cal M}_{2,k}^a
&=& -4\p^2 i~{\rm Tr} [\l_a^2 \g_a^{2k}]~ F^{aa}_k |_{\cal M}
\label{BGNDmobiusLAST} \eea
defining $F^{ab}_k$ as a term which contains all the
spin-structure and the orbifold information:
\be F^{ab}_k|_\s={1\over 4\p^4}\sum_{\a\b}\h_{\a\b}~ \p i
\partial_\t
\left[ \log{\vartheta[^\a_\b](0|\t)\over \h(\t)}\right]
{\vartheta[^\a_\b](0|\t)\over \h^3(\t)} Z^{ab}_{int,k}[^a_b]|_\s
\label{Fab-k}\ee
for both surfaces (the choice of $\t$ define the surface $\s$).
Note that the $a=b=1$ sector is not contained in the
(\ref{Fab-k}).
This term can be formally written as the supertrace over states
from the open $ab$ $k$-orbifold sector:
\be F^{ab}_k|_{\s}= {|G|\over (2\pi)^2}Str^{ab}_{k, {\rm
open}}\left[{1\over 12}-s^2\right]e^{-tM^2/2}\Big|_\s
\label{FkSTRACEI}\ee
where the $s$ is the 4D helicity.\\
Thus, for $\textit{large}$ $\t_2$ we have:
\be \lim_{\t_2\to\infty}F_k^{ab}=C_{k,IR}^{ab}+{\cal
O}[e^{-2\p\t_2}] \label{FonIR}\ee
with
\be C_{k,IR}^{ab}={|G|\over (2\pi)^2}Str_k \left[{1\over
12}-s^2\right]_{open} . \ee

For $\textit{small}$ $\t_2$ we have
\be \lim_{\t_2\to 0}F_k^{ab}={1\over \t_2}\left[C_{k,UV}^{ab}
+{\cal O}[e^{-{\p\over 2\t_2}}]\right]\label{FonUV}\ee
where
\be C_{k,UV}^{ab}={|G|\over (2\pi)^2}Str_k \left[{1\over
12}-s^2\right]_{closed} . \ee
The helicity supertrace is now in the closed-string $k$-sector
mapped from the open $k$-sector dy a modular transformation.

Notice that in the annulus amplitude (\ref{BGNDannulusLAST}), the
two first terms are proportional to the square of the $B$ field.
This cases are proportional to annulus amplitudes ${\cal A}_2$,
where two vertex-operators (VOs) are on the same boundary. In the
last component of (\ref{BGNDannulusLAST}), the $B$ fields are
coming from the opposite D-branes and is proportional to ${\cal
A}_{11}$, with the VOs on different boundaries. The
(\ref{BGNDmobiusLAST}) is proportional to a M\"obius strip
amplitude with the insertion of two VOs.

The IR limit $t \to \infty$ can be found easily using the
(\ref{FonIR}). We regularize the integral by $\m \to 1/t^2$ and we
find the $\b$-function:
\bea b &=&-{2\over N}\sum_{k=1}^{N-1}\lim_{t \to
\infty}\left({\cal A}_{2,k}^{ab}(t)+ {\cal
M}_{2,k}^{a}(t)\right)\nonumber\\
&=& -{2\p^2i\over N} \sum_{k=1}^{N-1} \bigg[\Big({\rm Tr} [\l_a^2
\g_a^k]{\rm Tr} [\g_b^k]+ {\rm Tr} [\g_a^k]{\rm Tr} [\l_b^2
\g_b^k] \nonumber\\
&& +2{\rm Tr} [\l_a\g_a^k]{\rm Tr} [\l_b\g_b^k] \Big)
C^{ab}_{k,IR}|_{\cal A} - 4 {\rm Tr} [\l_a^2 \g_a^{2k}]
C^{a}_{k,IR} |_{\cal M} \bigg] \eea
For the UV limit $t\to 0$, we use the (\ref{FonUV}) and we
regularize the integral by $\m \leq t$. The $A_2$ and $M$ together
are giving terms proportional to the tadpole cancellation
conditions\footnote{
The UV limit of $\partial_\t \log{\vartheta[^\a_\b] \over \h}$ in
(\ref{Fab-k}) is generically of order $\t_2^{-2}$. Terms that in
the closed sector appear as $\vartheta[^1_0]$ are contributions
from the RR part. These terms have limits $2\p i/ 3t^2$ and $\p i/
6t^2$ coming from the annulus and M\"obius strip respectively.
Terms that in the closed string sector appear as
$\vartheta[^0_\a]$ are the NSNS sectors which have contribution
only from the $\partial_\t\log\h$. The UV limits are $-\pi/3t^2$
and $-\pi/12t^2$ from the annulus and M\"obius strip respectively.
Therefore (\ref{BGNDannulusLAST}, \ref{BGNDmobiusLAST}) have the
same form as (\ref{BGNDannulus}, \ref{BGNDmobius}) that provides
the tadpole conditions. It is important that both, R and NS
sectors contribute to the mass formulas of the anomalous $U(1)$s.
%
}.
Therefore, when we have vanishing of RR and NSNS tadpoles, the
masses of the anomalous gauge bosons are given by ${\cal A}_{11}$:
\bea {1\over 2} M^2_{aa} &=&  {\p^2i \over N} \sum_{k=1}^{N-1}
{\rm Tr} [\l_a\g_a^k]^2
C^{ab}_{k,UV}|_{\cal A} \label{Mass-U1-aa}\\
{1\over 2} M^2_{59} &=&  {\p^2i \over 2 N} \sum_{k=1}^{N-1} {\rm
Tr} [\l_5\g_5^k]{\rm Tr} [\l_9\g_9^k] C^{59}_{k,UV}|_{\cal A}
\label{Mass-U1-59}\eea
where $\a=5,9$. When we have non-vanishing NSNS tadpoles there is
an extra contribution to the mass formulas, proportional to the
non-vanishing tadpole.

The formulae (\ref{Mass-U1-aa}, \ref{Mass-U1-59}) still hold even
if we add Wilson lines.
Generically, adding a Wilson line we shift the windings or the
momenta in a coordinate with Newmann or Dirichlet boundary
conditions respectively. This breaks the gauge group.
In the transverse (closed) channel the shifts appears as phases
$e^{2\p i n \q}$ where $\q$ the shift and $n$ the momenta or
windings respectively to the above.
Since only the massless states contribute in the UV limit, the
effect of the Wilson line will appear only in the traces of the
$\g$ matrices.

The threshold correction \cite{Kiritsis:1997hj} is the finite part
of (\ref{BGNDannulusLAST}) and (\ref{BGNDmobiusLAST}). Generically
we have:
\be {16 \p^2 \over g^2}={16 \p^2 \over g_0^2} -{1\over
2N}\sum_{k=1}^{N-1} \int^{1/\m^2}_\m {dt\over t}\left({\cal
A}_{2}^{ab}+ {\cal M}_{2}^{a}\right) - b \log{\m^2\over
M^2}-{1\over 2}M^2_{ab}{1\over \m}\ee
where we separate the divergencies from the quadratic terms to
$B$. The above formulae for the $\b$-function, the corrections to
the gauge couplings and the masses of the anomalous $U(1)$s are
the same to the supersymmetric ones found in
\cite{Antoniadis:2002cs, Bachas:bh}.
Next, we will apply the above formulae to a non-supersymmetric
model that has been constructed by Scherk-Schwarz deformation
\cite{Anastasopoulos:2003ha}.

\section{A 4d non-supersymmetric orientifold example}

In this section we will evaluate the masses of the anomalous
$U(1)$s in a $Z_2$ orientifold model where supersymmetry is broken
by a Scherk-Schwarz deformation
\cite{sss,kk,Antoniadis:1998ki,Cotrone:1999xs,Scrucca:2001ni,Ib1,Blum1}
and where RR and NSNS tadpoles cancel locally
\cite{Anastasopoulos:2003ha}. To start with, we give a review of
this model defining some useful quantities. Consider the ${\cal
N}=1$ orbifold of type IIB string theory in 4 dimensions,
$\Realint^4 \times T^2\times (T^4 /Z_2)$. The elements of this
orbifold are $\{1,g\}$, acting only on the $T^4$. In addition, we
can act with a freely-acting $Z_2$ orbifold with elements
$\{1,(-1)^F \d\}$. We denote by $h$ the non-trivial element of
this group. This orbifold is known as a Scherk-Schwarz
deformation. The $F=F_L+F_R$ is the space-time fermion number and
$\d$ is the element $(-1)^{m_4}$ (which geometrically corresponds
to the shift $x_4 \to x_4+\pi R_4$ of a compact dimension). As it
was shown in \cite{Anastasopoulos:2003ha}, the tadpole
cancellation provides two different solutions that depend on the
inequivalent choices of $\g_h^2=\pm 1$ where $\g_h$ the action of
$h$ on the Chan-Paton matrixes. The 16-dimensional 'shift' vector
of the $Z_2$ orientifold is \cite{Ibanez:1998qp,
Anastasopoulos:2003aj}:
\be V_g^9=V_g^5= {1\over 4}(1,1,1,1,1,1,1,1,1,1,1,1,1,1,1,1)~. \ee
The 'shift' vector of the SS deformation generically is:
\bea V_h^9=V_h^5={1\over 4} \left\{ \ba {ccc}
(1_a,-1_b) & \quad & \textrm{for $\g^2_h=-1$}\\
&&\\
(2_a,0_b)& \quad & \textrm{for $\g^2_h=+1$} \ea \right.
\label{V-h}\eea
where the index referred to the number of the same components in
the vector. In both cases $a+b=16$, however we implement for
simplicity $a=b=8$. The massless spectrums are provided in Table
1. The gauge group in both cases is the same. The only difference
appears in the exchange of the antisymmetric reps with the
bi-fundamental $(8,8)+(\overline{8},\overline{8})$ in the
(99)/(55) matter sector. The spectrum is anomaly-free in 4D since
it is non-chiral.
\begin{table}[h]
{\footnotesize \renewcommand{\arraystretch}{1.25}
\begin{tabular}{|c|c|c|}
\hline
\hline
$\g_h^2=-1$ & & \\
Gauge Group: $U(8)^2_9\times U(8)^2_{5}$ &
\raisebox{2.5ex}[0cm][0cm]{Scalars} &
\raisebox{2.5ex}[0cm][0cm]{Fermions}  \\
\hline
\hline
&  & $(28,1)+(\overline{28},1)+(1,28)$ \\
\raisebox{2.5ex}[0cm][0cm]{(99)/(55) matter} &
\raisebox{2.5ex}[0cm][0cm]{$(8,8)+ (\overline{8},\overline{8})$} &
$(1,\overline{28})+2\times(8;\overline{8})+2\times(\overline{8};8)$\\
%
\hline
& $(8,1;\overline{8},1)+ (\overline{8},1;8,1) $ &
$(8,1;1,\overline{8})+ (\overline{8},1;1,8)$ \\
\raisebox{2.5ex}[0cm][0cm]{(59) matter} & $(1,8;1,\overline{8})+
(1,\overline{8};1,8)$ & $(1,8;\overline{8},1)+
(1,\overline{8};8,1)$\\
\hline \hline
\hline
$\g_h^2=+1$ & & \\
Gauge Group: $U(8)^2_9\times U(8)^2_{5}$ &
\raisebox{2.5ex}[0cm][0cm]{Scalars} &
\raisebox{2.5ex}[0cm][0cm]{Fermions}  \\
\hline \hline  &  &   $(8,8)+
(\overline{8},\overline{8})$ \\
\raisebox{2.5ex}[0cm][0cm]{(99)/(55) matter} &
\raisebox{2.5ex}[0cm][0cm]{
$(28,1)+(\overline{28},1)+(1,28)+(1,\overline{28})$} & $2\times(8;
\overline{8})+2\times(\overline{8}; 8)$\\
\hline  & $(8,1;\overline{8},1)+ (\overline{8},1;8,1) $ &
$(8,1;1,\overline{8})+ (\overline{8},1;1,8)$ \\
\raisebox{2.5ex}[0cm][0cm]{(59) matter} & $(1,8;1,\overline{8})+
(1,\overline{8};1,8)$ & $(1,8;\overline{8},1)+
(1,\overline{8};8,1)$
\\
\hline
\end{tabular}}
\caption{The massless spectrum for the two inequivalent solutions
$\g_h^2=\pm 1$ of the $Z_2$ accompanied with a transverse SS
deformation. The gauge group in both cases is $U(8)_9\times
U(8)_{9'}\times U(8)_5 \times U(8)_{5'}$. The spectrum is
non-chiral and consequently anomaly-free.}
\end{table}
The internal annulus partition functions for 99, 55 and 59 strings
are:
\bea Z^{99,55}_{int,k}[^\a_\b] &=& - \sum_{s,r=0}^1 (-1)^{\a s +\b
r} \bigg[ (-1)^{s\cdot m_4}P_{m_4} P_{m_5} \bigg]
{\vartheta[^\a_\b](0|\t) \over \h(\t)} (2 \sin {\p k\over 2})^2
\prod_{j=1}^2 {\vartheta[^{~~\a}_{\b+2v_j k}](0|\t) \over
\vartheta[^{~~1}_{1+2v_j k}](0|\t)} \nonumber\\
Z^{59}_{int,k}[^\a_\b] &=& 2 \sum_{s,r=0}^1 (-1)^{\a s +\b r}
\bigg[(-1)^{s\cdot m_4}P_{m_4} P_{m_5}\bigg]
{\vartheta[^\a_\b](0|\t) \over \h(\t)} \prod_{j=1}^2
{\vartheta[^{~~\a+1}_{\b+2v_j k}](0|\t) \over
\vartheta[^{~~~0}_{1+2v_j k}](0|\t)}~. \label{Z-Internal}\eea
For $s=r=0$, we have the internal partition function of a
$T^2\times K^3 /Z_2$ orientifold. $s$ denotes the direct action of
the SS deformation and $r$ the twisted sector. The $(-1)^{s\cdot
m_4}P_{m_4} P_{m_5}$ is the lattice sum over momenta along the
first torus $T^2$:
\be (-1)^{s\cdot m_i} {P}_{m_i}({i\tau_2}/2) = {1
\over\eta({i\tau_2}/2)} \sum_{m_i} (-1)^{s\cdot m_i} q^{{\a'\over
4}\left({m_i\over R_i}\right)^2} \label{PI} \ee
For $s=1$ we have the SS deformation that shifts the $m_4$
momenta. As we mention before, $r=0,1$ denotes the $h$ untwisted
and twisted sectors respectively. However we will neglect the
twisted sector since it requires the insertion of anti-D-branes
\cite{Anastasopoulos:2003ha}.

To evaluate the masses of the anomalous bosons, we insert
(\ref{Z-Internal}) and (\ref{Fab-k}) in the mass
formulae\footnote{This model has local vanishing of RR and NSNS
tadpoles, and there will not be contributions from ${\cal A}_2$
and $\cal M$.}. After some 'thetacology' we find for $\a=5,9$:
\bea F^{\a\a}_{k=1} &=& {\h^2\over 2\p^2}\Big\{{\rm Tr} [\l^a\g_g]
{\rm Tr} [\l^a\g_g] + {\rm Tr} [\l^a\g_{gh}] {\rm Tr}
[\l^a\g_{gh}]~ (-1)^{m_4} \Big\} P_{m_4} P_{m_5}.
\label{F-Z2SS-aa}\\
F^{59}_{k=1} &=& -{\h^2\over 2\p^2}\Bigg\{{1\over 2}{\rm Tr}
[\l^5\g_g] {\rm Tr} [\l^9\g_g]\label{F-Z2SS-59}\\
&+&{i \over 2\p}{\vartheta^2_2 \vartheta^2_4 \over \h^6
\vartheta_3^2}
~\partial_\t \log {\vartheta_2 \vartheta_4 \over \h^2}~
{\rm Tr} [\l^5\g_{gh}]{\rm Tr} [\l^9\g_{gh}](-1)^{m_4}
\Bigg\}~P_{m_4} P_{m_5} . \nonumber\eea
The $\g$-matrices point out the sector that each term is coming
from. In the UV region, only the first terms in both formulae
contribute to the mass of the anomalous $U(1)$s. The second terms
(that contains the SS action $h$) after the Poisson re-summation
become proportional to $W_{\n_4+1/2}$ and does not contribute to
the $C^{99,55,59}_{UV}$. Since SS deformation does not contribute
to the mass terms of the anomalous $U(1)$s, we can directly
evaluate their masses for both two inequivalent solutions
($\g^2_h=\pm 1$):
\bea {1\over 2} M^2_{\a\a,ij} &=&  -{4 \p^2 \over 4} {\rm Tr}
[\l_i^a\g_g]{\rm Tr} [\l_j^a\g_g] {{\cal V}_1\over \p^2\a'}
\nonumber\\
&=&- {{\cal V}_1\over \a'} \left(-{i\over \sqrt{8}} \sin[2\p
V^a_i] \right) \left(-{i\over \sqrt{8}} \sin[2\p V^a_j] \right) =
{{\cal V}_1\over 8\a'}~.\label{Mass-U1-aa-Z2}\\
{1\over 2} M^2_{59,ij} &=&  {4 \p^2 \over 2\times 4} {\rm Tr}
[\l_i^5\g_g]{\rm Tr} [\l_j^9\g_g] {{\cal V}_1\over 2\p^2\a'}
= -{{\cal V}_1\over 32\a'}~.\label{Mass-U1-59-Z2}\eea
where $\a=5,9$. The mass-matrix has two massless gauge bosons
$-\tilde{A}_1+\tilde{A}_2$, $-A_1+A_2$ and two massive
$A_1+A_2+\tilde{A}_1+\tilde{A}_2$,
$-A_1-A_2+\tilde{A}_1+\tilde{A}_2$ with masses $3{\cal V}_1/
32\a'$, $5{\cal V}_1/ 32\a'$ respectively.

There are no anomalous $U(1)$s in these models since the spectrum
is non-chiral. However, the existence of the two massive gauge
bosons are the consequence of 6D anomalies \cite{Ibanez:2001nd,
Scrucca:2002is, Antoniadis:2002cs, Anastasopoulos:2003aj}. The
decompactification limit of the first torus (where the SS
deformation acts) leads to the N=1 6D $Z_2$ orientifolds that
contains two anomalous $U(1)$s that become massive via the
Green-Schwarz mechanism. Therefore, axions that participate in the
anomaly cancellation in the 6D model, contribute to the 4D masses
of the anomalous $U(1)$s by volume dependant terms. The ratio of
the masses found in \cite{Anastasopoulos:2003aj} for the $Z_2$
supersymmetric orientifold are the same to the above.

\section{Conclusion}

In this paper we evaluated the general mass formula for the
anomalous $U(1)$s in non-supersymmetric orientifolds. We have
shown that the supersymmetric formulae of \cite{Antoniadis:2002cs}
are also valid in non-supersymmetric orientifolds provided that
the tadpoles cancel.

Our analysis has direct implications for model building, both in
string theory and field theory orbifolds. It provides a necessary
and sufficient condition for a non-anomalous $U(1)$ to remain
massless (the hypercharge for example). The masses of the
anomalous $U(1)$s are always as heavy or lighter than the string
scale. Therefore, production of these new gauge bosons in particle
accelerators provides both constrains on model building and new
potential signals at colliders, if the string scale is around a
few TeV.

\newpage


\centerline{\bf\Large Acknowledgments} \vskip .5cm

The author would like to thank Elias Kiritsis for suggesting the
problem and discussions. He would like also to thank Constantin
Bachas, Amine B. Hammou, Angel M. Uranga for discussions. The
author was partially supported from RTN contracts
HPRN-CT-2000-00122 HPRN-CT-2000-0131 and INTAS contract
03-51-6346.

\bigskip\appendix
\section{Definitions and identities}

The Dedekind function is defined by the usual product formula
(with $q=e^{2\pi i\tau}$)
\be \eta(\tau) = q^{1\over 24} \prod_{n=1}^\infty (1-q^n)\ . \ee
The Jacobi $\vartheta$-functions with general characteristic and
arguments  are
\be \vartheta [^\a_\b] (z\vert\tau) = \sum_{n\in Z}
e^{i\pi\tau(n-\a/2)^2} e^{2\pi i(z- \b/2)(n-\a/2)}\ . \ee
We define:
$\vartheta_1(z|\t) = \vartheta \left[^1_1 \right] (z|\t)$,
$\vartheta_2(z|\t) = \vartheta \left[^1_0\right] (z|\t) $,
$\vartheta_3(z|\t) = \vartheta \left[^0_0\right] (z|\t) $,
$\vartheta_4(z|\t) = \vartheta \left[^0_1\right] (z|\t) $.
The modular properties of these functions are:
\bea \h(\t+1) = e^{i\p /12}\h(\t)\ , \ \ \vartheta
\left[^\a_\b\right] \left({z} \Bigl| {\t+1}\right)=
e^{-{i\pi\over4} \a(\a-2)}\vartheta \left[^{\ \ \ \a}_{\a
+\b-1}\right]
\left({z} \Bigl| {\t}\right) \nonumber\\
\h(-1/\t) = \sqrt{-i\t} \h(\t)\ , \ \ \vartheta
\left[^\a_\b\right] \left({z \over \t} \Bigl| {-1 \over
\t}\right)= \sqrt{-i \t} \ e^{i \p \left({\a \b\over 2} + {z^2
\over \t}\right)} \ \vartheta \left[^{\ \b}_{-\a}\right] (z | \t )
\label{f8} \eea
A very useful identity that is valid for $\sum h_i=\sum g_i=0$ is
\be \sum_{\alpha,\beta=0,1}\h_{\a\b}~
\vartheta\left[^\a_\b\right](v)\prod_{i=1}^{3}
\vartheta\left[^{\a+h_i}_{\beta+g_i}\right]
(0)=\vartheta_1(-v/2)\prod_{i=1}^{3} \vartheta
\left[^{1-h_i}_{1-g_i}\right](v/2)~. \label{SuperID}\ee

\end{document}